\documentclass[psfig]{aa} 

\usepackage{times}
\usepackage{graphics}
\usepackage{latexsym}
\usepackage{epsfig}

\begin{document}
%\headnote{Letter to the Editor}

\title{MAMBO 1.25\,mm observations of 3CR quasars at z$\sim$1.5: \\
  On the debate of the unified schemes} 

\author{M. Haas\inst{1}
   \and R. Chini\inst{1}
   \and S.A.H. M\"uller\inst{1}
   \and F. Bertoldi\inst{2}
   \and M. Albrecht\inst{3}
}
\offprints{Martin Haas (haas@astro.rub.de)}
\institute{
Astronomisches Institut, Ruhr-Universit\"at Bochum (AIRUB),
Universit\"atsstr. 150 / NA7, 44780 Bochum, Germany
\and
Radioastronomisches Institut, Universit\"at Bonn, Auf dem H\"ugel 71, 53121
Bonn, Germany
\and
Instituto de Astronom\'ia,
Universidad Cat\'olica del Norte,
Avenida Angamos 0610,
Antofagasta,
Chile
}
\date{Received 16. December 2004; accepted 02. September 2005 }
\authorrunning{M. Haas et al.}
\titlerunning{MAMBO observations of 3CR quasars}

\abstract{
  In order to explore the nature of the 850\,$\mu$m flux difference between
  powerful radio galaxies and steep radio-spectrum quasars
  at z$\sim$1.5 reported by Willott et
  al. (2002), we have observed 9 sources from their sample of 11 quasars at
  1.25\,mm. 
  For 7 sources the 1.25\,mm fluxes 
  are much brighter than one would expect from a purely thermal dust model
  fitted to the submm data,
  %lie clearly above
  %any reasonable thermal extrapolation from the 850\,$\mu$m data points, 
  providing
  evidence for the synchrotron nature of the observed 1.25\,mm radiation. 
  If we extrapolate a power-law synchrotron spectrum to
  shorter wavelengths, then for 6 of the 9 sources 
  also the 850\,$\mu$m fluxes are dominated by synchrotron
  radiation. 
  We discuss how far the (sub)-millimetre data can be interpreted 
  in accordance with the   
  orientation-dependent unified schemes for powerful 
  radio galaxies and quasars.
  In this case the results challenge the reported  evidence for the
  receding torus model and for 
  the evolutionary trend of a declining dust
  luminosity with increasing projected size of the radio lobes.

\keywords{Galaxies: fundamental parameters -- Galaxies: photometry -- Quasars: general -- Infrared: galaxies }
}
\maketitle

% ----------------------------------------------------------------
\section{Introduction} 
\label{section_introduction}

Here we consider two observational AGN classes,
the FR2 radio galaxies 
and the steep radio-spectrum quasars,
both with {\it powerful} edge-brightened radio lobes
(P$_{\rm 408 MHz}$ $\ga$ 10$^{\rm 24.5}$ W Hz$^{\rm -1}$).
According to the orientation-dependent unified schemes 
both classes are believed to 
belong to the same parent population with
the observed differences 
being just a consequence of their viewing angle: 
the jet axis of quasars is more aligned with our line of
sight than that of radio galaxies (Orr \& Browne 1982) and
the nuclear region of radio galaxies is hidden behind a dusty
torus seen roughly edge-on (Barthel 1989).
For these idealised versions of the unified schemes
two  predictions can be tested observationally: 
Firstly, at mm wavelengths the beamed synchrotron radiation should be higher
in the quasars than in the radio galaxies. 
Secondly, the isotropic far-infrared dust emission of
radio galaxies should be similar to that of quasars at matched
isotropic radio lobe power.

Both predictions have been confirmed
for 3CR sources at $z\la1.5$ by means of
sensitive ISO mid- and far-infrared and IRAM 
millimetre observations: 
For ten radio galaxy -- quasar pairs matched in 178\,MHz power 
Meisenheimer et al. (2001) found a similar dust detection
statistics as well as a higher synchrotron contribution in quasars. 
For a sample of eight sources van Bemmel \& Bertoldi (2001) inferred
also a higher synchrotron contribution in quasars. Using the full set
of ISO FIR observations of 75 sources Haas et al. (2004) found  
corroborating evidence for the orientation-dependent  unified schemes.

At redshift $z \ga 1.5$
the restframe FIR dust emission begins to shift 
into the submillimetre wavelength range.  
Based on SCUBA observations of 23 sources 
at $1.3<z<1.9$ Willott et al. (2002) found
that the quasars have a higher 850\,$\mu$m flux
than the radio galaxies.
They argue that the 850\,$\mu$m flux of these $z \sim 1.5$
quasars is dominated by dust
emission and not by synchrotron radiation. If this were true,
such a dust luminosity
difference between quasars and radio galaxies would not be consistent
with a simple version of the orientation-dependent unified schemes.
However, the thermal  
nature of the observed 850\,$\mu$m fluxes has still to be established.
In fact, the non-thermal synchrotron and free-free contribution in these sources is not
known. 
In the following we consider only thermal and synchrotron emission, 
since the data bases are too sparse to separate the observed fluxes
into three components and free-free emission generally plays a minor role in radio-loud
AGN; if significant, then free-free emission would not only reduce the synchrotron,
but also the thermal component. 

In order to explore the nature of the 850\,$\mu$m emission, 1.25\,mm
(or 3\,mm)
photometry allows to estimate
the synchrotron contribution at mm wavelengths
and to extrapolate it to the submm range.
Therefore, we observed the 
9 brightest quasars of Willott et al.'s sample, which
provided the highest likelihood of achieving a detection at 1.25\,mm with
the Max-Planck millimetre bolometer array MAMBO (Kreysa et al. 1998).

% ----------------------------------------------------------------
\begin{table*}
 \caption[] {Fluxes and other parameters of the quasar sample.
   The total submm fluxes are from Willott et al. (2002), observed
   between March 1999 and April 2001.
   Column 7 lists the thermal contribution at 1.25\,mm derived from a
   dust model fitted to the submm data, as explained in
   Sect.\,\ref{section_1250_nature} and shown in Fig.\,\ref{msxxxx_fig_seds_zoomed}.
   Column 8 lists the synchrotron contribution at 1.25\,mm, as total
   minus thermal contribution (col\,3 -- col\,7) and column 9 lists 
   the synchrotron fraction in percent. 
   The spectral indices $\alpha$$_{\rm cm-mm}$ are determined for most
   sources between 6\,cm and 1.25\,mm with an accuracy of better than 5\%;
   more details are given in Sect.\,\ref{section_850_nature}.
   These spectral indices are used to extrapolate from longer wavelength data  
   the 450 and 850\,$\mu$m synchrotron fluxes listed in columns 11 \& 12.
   \label{msxxxx_table1}
 }
  {\footnotesize
  \begin{center}
   \begin{tabular}{@{\hspace{0.5mm}}lr|rr|@{\hspace{0.5mm}}rr|@{\hspace{0.5mm}}rrr|@{\hspace{0.5mm}}c@{\hspace{0.5mm}}|@{\hspace{0.5mm}}rr@{\hspace{0.5mm}}}
~~~(1)&(2)~~~&(3)~~~&(4)~~~&(5)~~~&(6)~~~&(7)~~~&(8)~~~&(9)&(10)~~~&(11)~~~&(12)~~~\\
\hline                                                                                                                                                                               
  Object &   z      & F$_{1.25\,mm}$  & observ.& F$^{\rm total}_{\rm 850 \mu m}$ & F$^{\rm total}_{\rm 450 \mu m}$  & F$^{\rm thermal}_{\rm 1.25mm}$& F$^{\rm syn}_{\rm 1.25mm}$ & syn & $\alpha$$_{\rm cm-mm}$ & F$^{\rm syn}_{\rm 850 \mu m}$ & F$^{\rm syn}_{\rm 450 \mu m}$   \\
         &          &  [mJy]          & date     &    [mJy]          & [mJy]               &   [mJy]  &    [mJy]  & [\%] &                   & [mJy]                    & [mJy]                     \\
\hline                                                                                                                                                                               
  3C 181 & 1.382    & 5.96 $\pm$ 1.11 & 17.10.04 &   5.27 $\pm$ 1.06 &  $-$0.8 $\pm$ ~~7.5 &       0.66   &       5.30  & 88 &   $-$1.23            &   3.73                   &                           \\
  3C 191 & 1.952    & 5.36 $\pm$ 1.24 & 17.10.04 &   6.39 $\pm$ 1.06 &    28.9 $\pm$ ~~8.1 &       1.04   &       4.32  & 80 &   $-$1.17            &   3.44                   &   1.63                    \\
  3C 205 & 1.534    &                 &          &   2.36 $\pm$ 1.10 &    11.6 $\pm$ ~~7.9 &              &             &    &   $-$1.35\parbox{0cm}{$^{\rm b}$}& 2.24\parbox{0cm}{$^{\rm b}$}&            \\
3C 268.4 & 1.400    & 8.53 $\pm$ 1.56 & 16.10.04 &   5.13 $\pm$ 1.18 &     9.9 $\pm$  12.5 &       1.11   &       7.42  & 86 &   $-$1.10            &   5.55                   &                           \\
3C 270.1 & 1.519    &11.92 $\pm$ 2.69 & 02.06.04 &   7.42 $\pm$ 1.17 &    15.4 $\pm$  11.5 &       1.07   &      10.85  & 91 &   $-$1.11            &   7.76                   &   3.83                    \\
3C 280.1 & 1.659    & 3.99 $\pm$ 1.24 & 16.10.04 &   5.10 $\pm$ 1.74 &     0.4 $\pm$  23.8 & $<$1.31\parbox{0cm}{$^{\rm a}$}&$>$2.68\parbox{0cm}{$^{\rm a}$}& 67 &  $-$1.15            &   2.62                   &                           \\
3C 298   & 1.439    &20.68 $\pm$ 3.68 & 02.06.04 &  21.13 $\pm$ 2.06 &  $-$8.0 $\pm$  23.7 &       2.12   &      18.56  & 89 &   $-$1.10            &  13.88                   &                           \\
3C 318   & 1.574    & 5.61 $\pm$ 1.42 & 02.06.04 &   7.78 $\pm$ 1.00 &    21.6 $\pm$  10.6 &       0.99   &       4.62  & 82 &   $-$1.32            &   3.53                   &                           \\
3C 432   & 1.785    & 0.34 $\pm$ 1.54 & 29.05.04 &   7.93 $\pm$ 1.70 &     2.9 $\pm$  20.9 &       2.06   &   $<$ 2.56  & 55 &   $-$1.25\parbox{0cm}{$^{\rm b}$}& 1.60\parbox{0cm}{$^{\rm b}$}&            \\
4C 38.30 & 1.405    &                 &          &$-$0.01 $\pm$ 1.11 &     8.9 $\pm$ ~~7.8 &              &             &    &   $-$1.10\parbox{0cm}{$^{\rm b}$}& 1.14\parbox{0cm}{$^{\rm b}$}&            \\
4C 35.23 & 1.594    &11.93 $\pm$ 2.89 & 02.06.04 &   9.18 $\pm$ 1.10 &    21.3 $\pm$ ~~6.8 &       0.68   &      11.25  & 94 &   $-$0.89            &   8.60                   &   4.88                    \\     
     \hline
   \end{tabular}
  \end{center}
$^{\rm a}$ upper and lower limits adopted, because the thermal model is constrained by submm upper flux limits only.\\
$^{\rm b}$ determined/extrapolated from the cm wavelength range only.
  }
\end{table*}
%88
%80
%86
%91
%67
%89
%82
%55
%94
%----------------------------------------------------------------------------------------------------------

\section{Observations and Data}
\label{section_observations}
The MAMBO 1.25\,mm (250 GHz) continuum observations were performed
at the IRAM 30-m telescope during the
pooled observation campaigns between May and October 2004.  We
used the standard on-off photometry observing mode, chopping between
the target and sky at 2 Hz, and nodding the telescope every 10 s.
The atmospheric transmission was intermediate with
$\tau$(1.2\,mm) between 0.15 and 0.5. 
The absolute flux calibration was established by observations of
Mars and Uranus, resulting in a flux calibration uncertainty of
about 20\%.  The data were reduced using the MOPSI software package.

The 1.25\,mm fluxes are listed in Table\,\ref{msxxxx_table1},
together with the submm fluxes and other parameters discussed below.
Figure\,\ref{msxxxx_fig_seds} shows the spectral energy distributions
(SEDs), with the submm-mm range being zoomed in
Fig.\,\ref{msxxxx_fig_seds_zoomed}.
At a redshift $z \ga 1.5$ the host galaxies are unresolved. 
The 1.25\,mm fluxes refer to a beam size of 11$\arcsec$ of the
IRAM 30\,m telescope, which is similar to the JCMT-SCUBA 850\,$\mu$m beam 
of 15$\arcsec$, but smaller than the size of the extended radio
structures measured at cm wavelengths. 
Any extended contribution to the submm and mm fluxes, if
significant at all, may be missed, but we will see below
(Sect.\,\ref{section_850_nature}) that this effect might be negligible.

\section{Results and Discussion}
\label{section_results}

We investigate our quasar 1.25\,mm data and their implications 
on the nature of the 850\,$\mu$m emission. In order to test the unified
schemes, we then compare the quasars with radio galaxies. 

\subsection{Quasars}
\label{section_quasars}

This section is sub-divided into three steps:
(1) in order to determine the nature of the 1.25\,mm fluxes
we firstly consider the extreme case adopting 
that the submm fluxes are entirely thermal;
(2) since we find that for most sources the 1.25\,mm fluxes are dominated by
synchrotron radiation, we explore its influence on the nature
of the 850\,$\mu$m fluxes;
(3) finally we consider the remaining evidence for thermal 850\,$\mu$m
emission. 

\subsubsection{Synchrotron nature of the 1.25\,mm fluxes}
\label{section_1250_nature}

The most striking result from Fig.\,\ref{msxxxx_fig_seds_zoomed}
is the strong evidence that the 1.25\,mm data
are much brighter than one would expect from a purely thermal dust model
fitted to the submm data.

For the first step we assume that the
submm fluxes are entirely due to dust emission.
With a dust emissivity $\beta$ = 2 a
greybody fit to the 450 and 850\,$\mu$m data yields a dust
temperature in the range 40\,K\,$<$\,T\,$<$\,60\,K
for all sources, except 4C35.23 discussed below.
Because of the numerous 450\,$\mu$m upper limits 
the individual temperature values are quite uncertain.
Therefore, for all sources we use 
T\,=\,50\,K typically found for quasars with high far-infrared
and submm luminosity  above 10$^{\rm 12}$ L$_{\odot}$
(e.g. Willott et al. 2002, Haas et al. 2003, 2004).
Fig.\,\ref{msxxxx_fig_seds_zoomed} shows the SEDs zoomed around the
submm-mm data and a modified
blackbody at T\,=\,50\,K; the strength of this greybody is constrained by the
450\,$\mu$m data points for all sources except 3C280.1, for which the
450\,$\mu$m upper limit is rather high and we used the 850\,$\mu$m
constraint. 

For 4C35.23 the greybody fit to the 450 and 850\,$\mu$m data
results in a low dust temperature T $\approx$ 20K;
since this is an extremely unusual value for such
an active high-luminosity object,
and since the 850\,$\mu$m data point lies nicely on the
synchrotron extrapolation
(as shown in Fig.\,\ref{msxxxx_fig_seds_zoomed} and discussed further below),
this argues in favour of the warmer, say T\,$\approx$\,50\,K, dust component
running through the
450\,$\mu$m data point only.
Even more extreme, the attempt to fit the 850\,$\mu$m and 1.25\,mm data
by a greybody results in exceptionally low temperatures, for
example T\,$\approx$\,9\,K for 3C191 and 3C318, and even much lower for the
other sources (except 3C432). Since we hesitate to postulate the
existence of such a cold bright dust component in luminous quasars,
we conclude that the 1.25\,mm fluxes are mainly due to 
synchrotron radiation.
Table\,\ref{msxxxx_table1} lists the thermal 1.25\,mm contribution derived
from the T\,=\,50\,K blackbody shown in Fig.\,\ref{msxxxx_fig_seds_zoomed},
as well as the 1.25\,mm synchrotron contribution as difference of
total minus thermal 1.25\,mm flux.
For at least seven quasars (i.e. all except 3C\,280.1 and 3C\,432) the 
synchrotron fraction lies above $\sim$80\% of the total 1.25\,mm
flux.
Considering the errors, if we adopt the extreme case that all actual
1.25\,mm fluxes are 20\% lower, then
for seven sources the synchrotron fraction still remains above $\sim$75\%.
This fraction becomes higher, if we do not assume that the
submm fluxes are of thermal nature. 

% ----------------------------------------------------------------
\begin{figure*}
  \hspace{-0.25cm}
  \epsfig{file=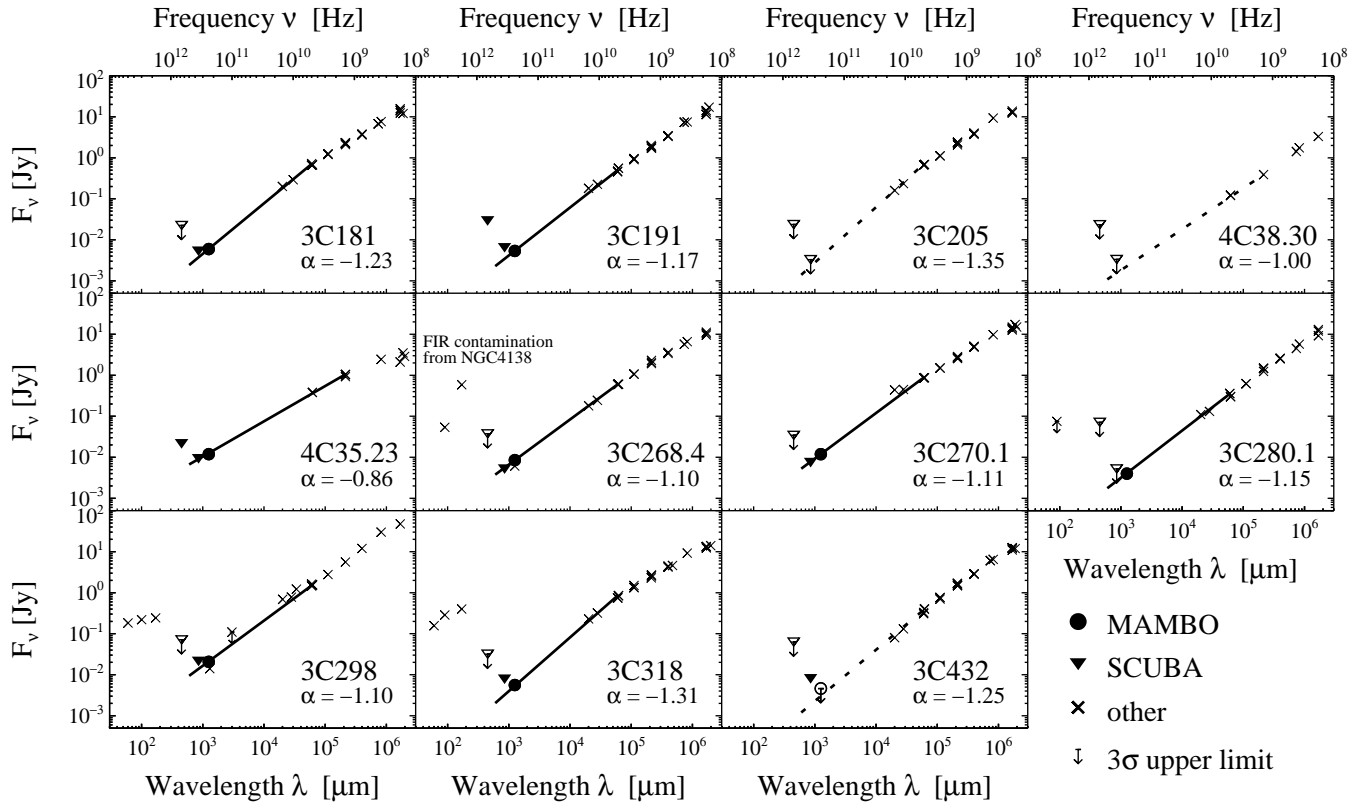,width=18.5cm, clip=true}
  \caption[]{ \label{msxxxx_fig_seds} 
    Observed SEDs of the quasars.
    The errors are smaller than the size of the symbols;
    open symbols with arrows represent 3$\sigma$ upper limits.
    The drawn straight lines show the synchrotron spectrum from the cm to the mm wavelength range
    with spectral index $\alpha$ (=$\alpha$$_{\rm cm-mm}$ in Tab.\,\ref{msxxxx_table1});
    these lines are dotted in case of an upper limit
    or a missing data point at 1.25\,mm.
    As already noted by Haas et al. (2004)
    the exceptionally high FIR 90 and 170$\mu$m fluxes of 3C268.4
    reported by Andreani et al. (2002)
    are contaminated by the nearby galaxy NGC4138.
  }
\end{figure*}
% ----------------------------------------------------------------

% ----------------------------------------------------------------
\begin{figure*}
  \hspace{-0.25cm}
  \epsfig{file=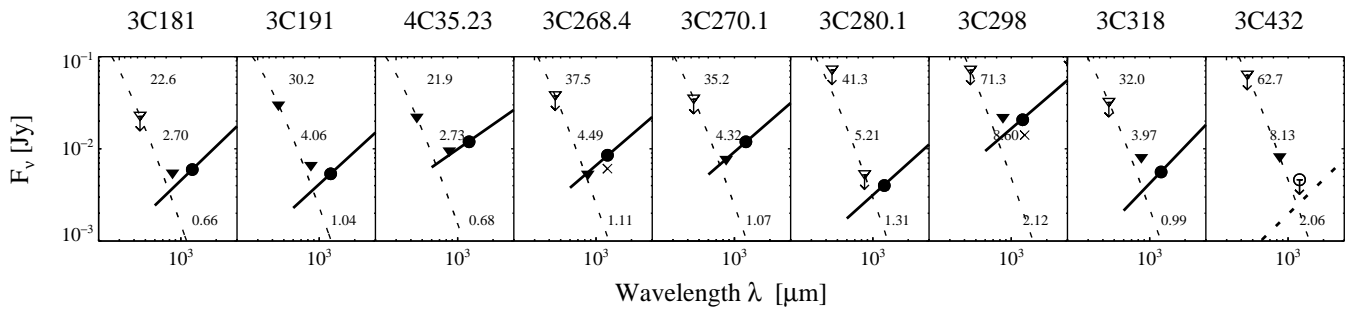 ,width=18.5cm, clip=true}
  \caption[]{ \label{msxxxx_fig_seds_zoomed} 
    SEDs of the nine quasars observed at 1.25 mm, zoomed around the
    submm-mm data points. The errors are in the order of the size of
    the symbols. 
    The symbols are as in Fig.\,\ref{msxxxx_fig_seds}.
    The dotted lines show a 50 K modified blackbody
    (emissivity $\beta$=2) with three numbers in each panel listing its  
    flux values at 450, 850 and 1250\,$\mu$m ;
    the strength of this greybody is constrained by the
    450\,$\mu$m data for all sources except for 3C280.1 for which it is
    constrained by the 850\,$\mu$m upper limit.
    The plots clearly illustrate the non-thermal nature of the 1.25\,mm fluxes 
    for at least seven sources, the exceptions are 3C\,432 and perhaps also 3C\,280.1.
    The drawn straight lines show the synchrotron
    spectrum as in Fig.\,\ref{msxxxx_fig_seds}. 
    }
\end{figure*}
% ----------------------------------------------------------------

In the discussion here, we do not make use of a possible synchrotron  
variability between different observing dates,
since suitable monitoring data are not available for the entire sample.
Two sources, notably, have been observed also at 1.25\,mm on other dates:
3C\,298 (14.1 $\pm$ 0.8 mJy in Dec. 1998, Meisenheimer et al. 2001)
and 
3C\,268.4 (6.1 $\pm$ 1.0 in March 1996, Andreani et al. 2002). Compared with
our data taken in summer/autumn 2004 (Tab.\,\ref{msxxxx_table1}) 
some variability may be present, for 3C\,298  also indicated by 
the irregular deviation of the cm flux values from a
straight line
(Fig.\,2 in Willott et al. 2002).
In these two cases, however, any variability appears to be moderate ($\la$30\%)
and lies within the calibration and measurement errors.

\subsubsection{Synchrotron contribution at 850\,$\mu$m}
\label{section_850_nature}

The high 1.25\,mm synchrotron fraction 
suggest that also the 850\,$\mu$m fluxes
may contain a high synchrotron contribution.
Notably, in this case our initial assumption
on the purely thermal
nature of the submm fluxes would be invalidated, further
reinforcing the synchrotron nature of the 1.25\,mm emission
concluded above.

The precise extrapolation of the 1.25\,mm synchrotron contribution to
shorter wavelengths depends on the shape and slope of the synchrotron
spectrum. Thereby two aspects are important:

%\begin{itemize}
%\item[1)] 
1) As discussed by Willott et al. (2002) and references therein
a study of several steep radio-spectrum sources between
22\,GHz and 230\,GHz shows that the radio {\it cores} are the essential
contributors to the 1.25 mm and 850\,$\mu$m synchrotron fluxes. The cores  
have a typical spectrum F$_\nu$\,$\propto$\,$\nu$$^{\alpha}$  with, on average, 
$\alpha$\,$\sim$\,$-1$. 
Applying this average $\alpha$ value for the extrapolation of the 1.25\,mm data
to shorter wavelengths clearly results in high 850\,$\mu$m synchrotron
contributions which, however, could be caused by the scatter of
$\alpha$. Therefore we try to
obtain better estimates for the individual sources.
Due to the lack of sufficient
radio {\it core} data of our sources we here investigate their {\it total}
spectra. 

%\item[2)] 
2) In principle, there
may be a spectral break at wavelengths between 1.25\,mm and
850\,$\mu$m (for both core and total spectra).
To our knowledge, however, such an extreme break, occurring within a factor of
only 1.5 in frequency range, has not yet been observed so far.
Also, according to theoretical models such a break should show
up already via a deviation from the power-law shape
extending over at least a factor of ten in frequency, i.e. 
a pronounced downward curvature of the spectrum between cm and mm
wavelengths
(e.g. Pacholzyk 1980, see Fig.\,3 in Marscher 1977 and Fig.\,1 in 
Crusius \& Schlickeiser 1986). 
In order to check for such a curvature, additional 3\,mm data would be ideal,
but they are not available for our sources. Therefore, we here
investigate the present cm-mm spectra. 

%\end{itemize}

The aim now is to check for signatures of a curvature between cm and mm
wavelengths, which would indicate a spectral break
between 1.25\,mm and 850\,$\mu$m 
and, if such signatures are not found, to get an estimate on the
850\,$\mu$m synchrotron contribution via power-law extrapolation from
the 1.25\,mm data points. 

In order to facilitate the following analysis we start with the assumption 
that the 1.25\,mm fluxes are entirely due to synchrotron emission. 
In a first step we visually fitted the spectral indices
$\alpha$ at wavelengths 2\,cm $\la$ $\lambda$ $\la$ 21\,cm.
Drawing these lines further down to mm wavelenths
results in extrapolated 1.25\,mm fluxes close to
those observed. In a refined second step we fitted the spectral
indices including also the 1.25 mm data. 
The exact wavelength range used depends on the available data.
In general $\alpha$$_{\rm cm-mm}$ was determined between 6\,cm and 1.25\,mm;
slightly different wavelength ranges were used for 3C205 (6\,cm\,-\,2\,cm),  
4C38.30 (21\,cm\,-\,2\,cm),
and 4C35.23 (21\,cm\,-\,1.25\,mm),
yielding $\alpha$$_{\rm cm-mm}$ values similar to
those of the other sources.
The spectral indices $\alpha$$_{\rm cm-mm}$ lie between $-0.9$ and $-1.35$, a
range also found for quasars and BLRGs by van Bemmel \& Bertoldi
(2001). While we had to choose slightly different cm wavelength
ranges, the inclusion of the 1.25\,mm data point makes the fit
of $\alpha$$_{\rm cm-mm}$ very stable with uncertainties below 2\%.
For the two cases without mm data we used the shortest cm spectral range
which follows a
power-law, and we estimate the formal fit uncertainty of  
$\alpha$ to be less than 5\% which is negligibly small compared with
the influence of other effects like variability or a possible spectral
curvature between cm and mm  wavelengths. 

The fact that the 1.25\,mm data points lie on (or very close to) the
power-law extrapolation of the radio cm 
spectra to shorter wavelengths (Figure\,\ref{msxxxx_fig_seds})
suggests that any spectral curvature around 1\,mm is weak, hence that 
there is no strong spectral break between 1.25\,mm and 850\,$\mu$m. 
Also, it indicates that probably 
not much extended mm flux is missed as
mentioned above (Sect.\,\ref{section_observations}), thereby 
arguing in favour of the synchrotron nature of the 1.25\,mm fluxes, too.  
Both results are also valid, if we modify the assumption that (instead
of 100\%) only 75-80\%
of the 1.25\,mm fluxes are synchrotron radiation. 

Since the synchrotron spectra follow the
F$_\nu$$\propto$$\nu$$^{\alpha}$ extrapolation from the cm
fluxes to shorter wavelengths and show neither any hint for an
abrupt break nor a strong curvature, we suggest that the synchrotron
contribution at 850\,$\mu$m can be extrapolated reasonably well 
from the 1.25\,mm data points using
these spectral indices $\alpha$$_{\rm cm-mm}$. 
Noteworthy, $\alpha$$_{\rm cm-mm}$ determined from the total spectra
is steeper than the average core value $\alpha$\,$\sim$\,$-1$ (except for 4C\,35.23);
this suggests that a spectral break between 1.25\,mm and 850\,$\mu$m, if any,
is even weaker.
Therefore, 
compared with using $\alpha$\,$\sim$\,$-1$ mentioned above, one may expect that 
the total $\alpha$$_{\rm cm-mm}$
yields a lower than actual 850\,$\mu$m synchrotron contribution. 
This contribution (Tab.\,\ref{msxxxx_table1}) already 
constitutes at least 50\% of the observed total 
850\,$\mu$m flux for five sources (3C\,181, 3C\,191, 3C\,280.1, 3C\,298, 3C\,318)
and reaches about 100\% for three sources
(3C\,268.4, 3C\,270.1, 4C\,35.23);
only for one source (3C\,432) it lies at about 20\%.
Also for the two sources without 1.25\,mm data, the tentative extrapolation of the
cm spectra yields a 850\,$\mu$m synchrotron fraction of $\ga$30\% (4C\,38.30)
and 100\% (3C\,205).
Thus, as long as future data do not reveal a synchrotron break between
1.25\,mm and 850\,$\mu$m, we conclude that 
in most (9/11) sources not only the 1.25\,mm fluxes,
but also the 850\,$\mu$m fluxes are dominated by synchrotron
radiation.

% ----------------------------------------------------------------
\begin{figure*}
  \hspace{-0.25cm}
  \epsfig{file=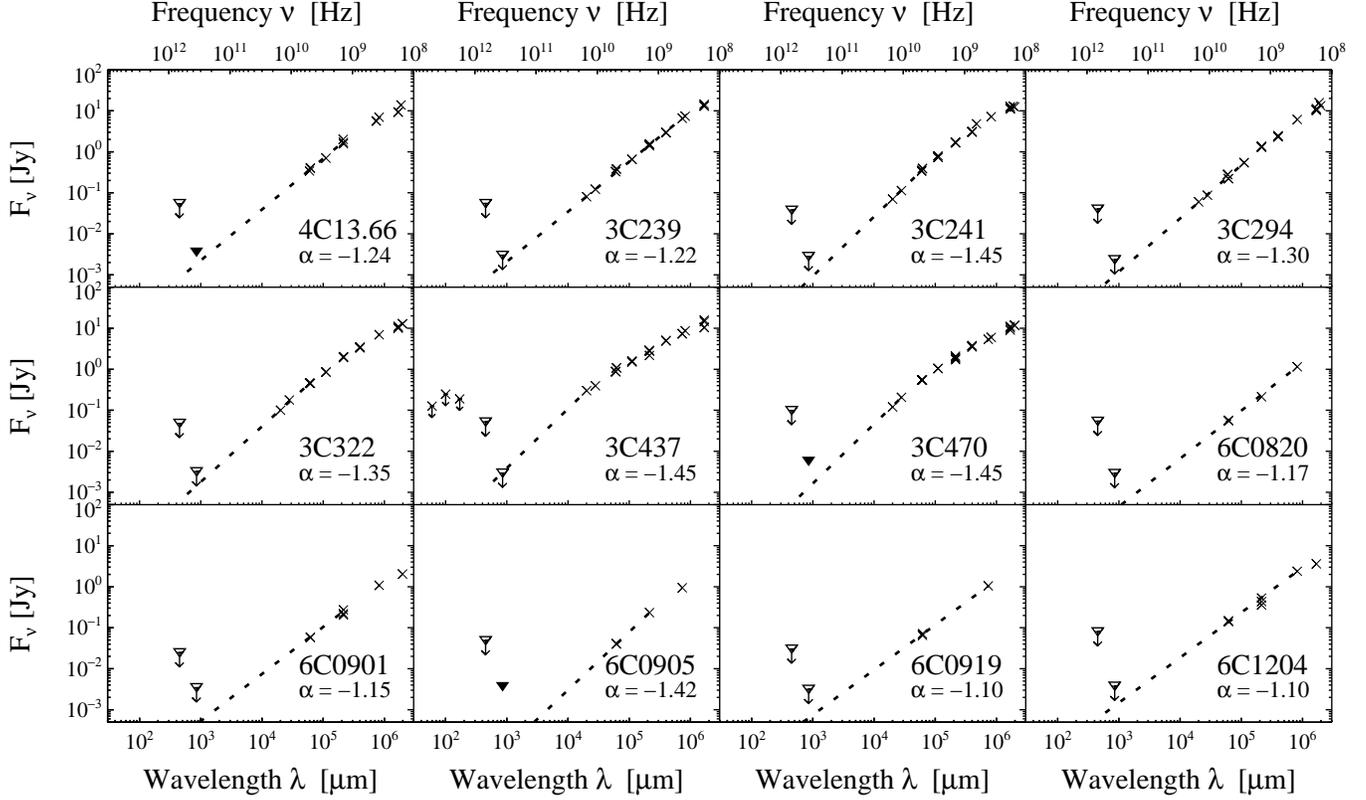 ,width=18.5cm, clip=true}
  \caption[]{ \label{msxxxx_fig_gal_seds} 
    Observed SEDs of the comparison radio galaxies.
    Symbols are as in Fig.\,\ref{msxxxx_fig_seds}.
    }
\end{figure*}
% ----------------------------------------------------------------

\subsubsection{Thermal contribution at 850\,$\mu$m}
\label{section_thermal}

Although one could fit a combined model consisting of a thermal and a
synchrotron component to the SEDs, we do not present nice fits
here because they do not provide further insight and bear the risk
of diluting the attention for the implicit assumptions made; instead we
prefer to extract the conclusions from considering various cases with
clear assumptions.

Again, we assume in a first step a 100\% synchrotron nature of the 1.25\,mm
fluxes and extrapolate the 850\,$\mu$m synchrotron contribution using
$\alpha$$_{\rm cm-mm}$ derived from the total
sources.
In order to determine which sources still
show evidence for thermal 850\,$\mu$m emission,  
from the entire 850\,$\mu$m fluxes
we subtract the synchrotron contribution. 

From inspection of Figs.\,\ref{msxxxx_fig_seds} and
\ref{msxxxx_fig_seds_zoomed} and with values listed in
Table\,\ref{msxxxx_table1} we find only five such quasars out of eleven.
Three show evidence for thermal submm emission from data at both 450 and
850\,$\mu$m (3C\,191, 3C\,318, 3C\,432) ,
one source (4C\,35.23) only from data at 450\,$\mu$m,
and for one source (3C\,280.1) the rather high upper submm flux limits may
allow for thermal 850\,$\mu$m emission. 
For the other four quasars any evidence for thermal 850\,$\mu$m
emission remains weak, even if we reduce the 1.25\,mm synchrotron
fraction from 100\% down to 75\%; in order to allow
for a dominant thermal 850\,$\mu$m emission in these sources a strong spectral
break between 1.25\,mm 850\,$\mu$m would be required, but the current
data do not show signatures in favour of such a break.  

In the two cases without 1.25\,mm data (3C\,205 and 4C\,38.30), the observed 
850\,$\mu$m upper limits lie only marginally above the synchrotron spectrum 
extrapolated from the cm wavelengths; thus, 
considering the entire sample this would result
in six out of eleven quasars showing no evidence for a significant
thermal 850\,$\mu$m emission.

\subsection{Comparison with radio galaxies}
\label{section_comparison_with_radio_galaxies}

In order to test the unified schemes we compare
the quasars with 
the sample of radio galaxies at matched redshift and
151\,MHz radio lobe power 
selected by Willott et al. (2002) and observed by Archibald et al. (2001).

\subsubsection{Properties of the radio galaxies}
\label{section_prop_gal}

Figure\,\ref{msxxxx_fig_gal_seds} shows the SEDs of the radio
galaxies, from the submm to the cm wavelength range, referring to the total
fluxes. Suitable mm data are not 
available for a direct determination of the synchrotron
contribution at submm-mm wavelengths.
However, we will see below that basic conclusions can be drawn even
without such mm data. 

For the quasars the power-law fit to the total spectra at
cm wavelengths yields synchrotron slopes with values close to
those $\alpha$$_{\rm cm-mm}$ obtained including also the 1.25\,mm data
to the fit.
Encouraged by this fact
we tentatively determine $\alpha$ in the cm regime for the radio galaxies 
and consider its extrapolation to the mm and submm wavelengths
(Fig.\,\ref{msxxxx_fig_gal_seds}). As for the quasars, for the radio galaxies 
we visually fitted the spectral indices
$\alpha$ at wavelengths 21\,cm $\ge$ $\lambda$ $\ge$ 2\,cm.
Depending on the available data
slightly different wavelength ranges were used 
for 
4C\,13.66 and the 6C radio galaxies (21\,cm\,-\,6\,cm), and 
for most (4/6) of the 3C galaxies  (21\,cm\,-\,2\,cm).   
Since two 3C galaxies (3C\,437 and 3C\,470) show a slight curvature in their cm
spectra, i.e. a steepening of the spectra towards
shorter wavelengths, 
we used only the shortest cm range (6\,cm\,-\,2\,cm) to determine
their $\alpha$. 
In one case (3C\,437) the 850\,$\mu$m data point indicates a 
synchrotron spectrum which is even steeper, 
and we used this steep $\alpha$ value constrained by 
the 850\,$\mu$m data point. 
Table\,\ref{msxxxx_table2} lists the values for $\alpha$$_{\rm cm}$ and the
extrapolated synchrotron contribution at 850\,$\mu$m.
Because some radio galaxies (e.g. 3C\,437 and 3C\,470) show the trend of a
convex curvature in their cm spectra, for the others 
one may expect that the $\alpha$ values determined from the
cm range actually represent upper limits for the true 
$\alpha$$_{\rm cm-mm}$; 
in this case the actual  850\,$\mu$m synchrotron
contributions will be even lower than the extrapolated values.

% ----------------------------------------------------------------
\begin{table}
  \begin{center}
 \caption[] {Parameters of the comparison radio galaxy sample of Willott et al. (2002): 
   Spectral indices, $\alpha$$_{\rm cm}$, fitted to the total cm wavelengh
   data with an accuracy of about 5\%.
   Via F$_\nu$\,$\propto$\,$\nu$$^{\alpha}$ the
   850\,$\mu$m synchrotron fluxes are extrapolated from the cm 
   fluxes with a formal uncertainty of about 30-40\%.
   \label{msxxxx_table2}
   }
  {
   \begin{tabular}{lc|cc}
  Object    &   z   & $\alpha$$_{\rm cm}$& F$^{\rm syn}_{\rm 850 \mu m}$ [mJy] \\
%            &       &                     &                   \\
\hline                                                            
4C13.66     & 1.450 &   $-$1.24           &  1.84                   \\
3C239       & 1.781 &   $-$1.22           &  1.71                   \\
3C241       & 1.617 &   $-$1.45           &  0.73                   \\
3C294       & 1.786 &   $-$1.30           &  0.96                   \\
3C322       & 1.681 &   $-$1.35           &  1.44                   \\
3C437       & 1.480 &   $-$1.45\parbox{0cm}{$^{\rm a}$}&  3.05                   \\
3C470       & 1.653 &   $-$1.45           &  1.30                   \\
6C0820+3642 & 1.860 &   $-$1.17           &  0.37                   \\
6C0901+3551 & 1.904 &   $-$1.15           &  0.43                   \\
6C0905+3955 & 1.882 &   $-$1.42           &  0.09                   \\
6C0919+3806 & 1.650 &   $-$1.10           &  0.62                   \\
6C1204+3708 & 1.779 &   $-$1.10           &  1.24                   \\
     \hline
   \end{tabular}
  }

$^{\rm a}$ actually $\alpha$$_{\rm cm}$ refers to $\alpha$$_{\rm cm-850 \mu m}$, \\
due to the low 850\,$\mu$m upper limit
\end{center}
\end{table}
% ----------------------------------------------------------------

Two radio galaxies 
(3C\,470 and 6C0905) and possibly also a third one 
(4C\,13.66) have a detected total 850\,$\mu$m
flux lying above this synchrotron extrapolation
(Fig.\,\ref{msxxxx_fig_gal_seds}, Table\,\ref{msxxxx_table2}), 
hence show signatures for 
thermal submm emission. Furthermore, in three radio galaxies 
(6C\,0820, 6C\,0901 and 6C\,0919) 
the 850\,$\mu$m upper flux limits lie 
clearly above the synchrotron extrapolation, 
thus in principle allowing for a significant 
thermal submm contribution, too. 

One may ask how far additional 1.25\,mm 
data for the radio galaxies could change this picture.
We consider two basic cases assuming that variability, if any, is
small:

1) If the mm data point would lie so much above the current extrapolation that
the evidence for thermal 850\,$\mu$m emission disappears 
(e.g. for 3C\,470, 6C\,0905), then the
synchrotron spectra of the total sources 
would have a strong upward curvature between cm and mm wavelengths.
But in the total spectra of radio galaxies such a strong mm bump 
has not been observed so far. 
Also, 
this possibility was discussed in detail by Archibald et al. (2001, their
Sect.\,4.2) for those sources in their larger (z=0.7-4.4) 
radio galaxy sample which have radio core data. 
Even for the worst-case scenario
any synchrotron contribution to the
850\,$\mu$m core fluxes is negliglble ($\la$\,0.2\,mJy);
an exception may be 3C\,241 likely to have 
F$^{\rm syn}_{\rm 850 \mu m}$\,$\sim$\,1\,mJy, a value  consistent with our
extrapolation of  F$^{\rm syn}_{\rm 850 \mu m}$\,$\sim$\,0.73\,mJy
(Table\,\ref{msxxxx_table2}).

2) If the mm data point lies below the current extrapolation, then the
synchrotron contribution might be even lower
and the thermal component higher than derived by our current procedure.

These two considerations suggest that  1.25\,mm data are 
not necessary for drawing the relevant conclusions here.
Actually the reason for not obtaining 1.25\,mm data for the radio galaxies (and
the two faintes quasars) was that  at the expected low
flux levels F$_{\rm 1.25 mm}$\,$\la$\,1\,mJy such observations are very time
consuming ($\ga$\,10 hours per object).

% ----------------------------------------------------------------
\begin{figure}
  \epsfig{file=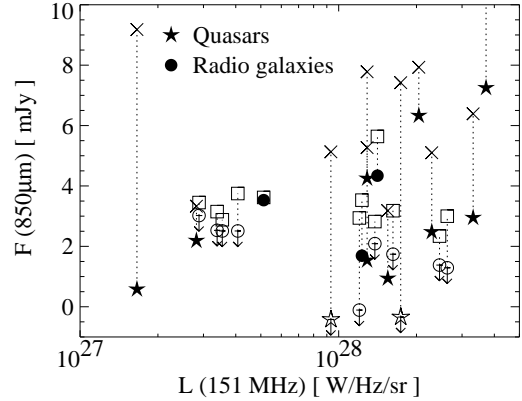 ,width=8cm, clip=true}   
  \caption[]{ \label{msxxxx_fig_distributions} 
    Distribution of 850\,$\mu$m fluxes versus radio lobe luminosities.
    For each object the total flux and -- after subtraction of the
    synchrotron contribution -- the thermal flux is shown, connected by a
    vertical dotted line.   
    Open symbols with arrows indicate 3-$\sigma$ upper limits.
    A proper thermal comparison would be via dust luminosities at a given restframe
    wavelength, but would require more data or assumptions on the dust
    temperature.
    }
   \vspace{-0.2cm}   
\end{figure}
% ----------------------------------------------------------------

\subsubsection{Comparison of the distributions}
\label{section_distributions}

The 850\,$\mu$m synchrotron contribution for the
quasars lies in a range with an average of 4.92$\pm$3.82 mJy, which is
about a factor of 5 higher
and shows a broader distribution than that for radio galaxies
(1.15$\pm$0.81 mJy). The distributions of spectral slopes $\alpha$$_{\rm cm-mm}$
for quasars and $\alpha$$_{\rm cm}$ for radio galaxies 
overlap. The average slopes   
are $-1.16$$\pm$0.13 and $-1.28$$\pm$0.14, resp.,
hence are slightly steeper for the radio galaxies.
Both results are  consistent with the
orientation-dependent unified schemes, where the jet axis of the quasars
is thought to be more aligned with our line of sight. 

The 850\,$\mu$m dust contributions F$^{\rm dust}_{\rm 850 \mu m}$ 
after subtraction of the synchrotron contribution
from the total fluxes are 
shown in Fig.\,\ref{msxxxx_fig_distributions}.
For three quasars and two radio galaxies 
F$^{\rm dust}_{\rm 850 \mu m}$ lies above 3\,mJy,
a value we adopt as observational 3-$\sigma$ detection threshold 
(Archibald et al. 2001).
For six quasars and one radio galaxy, which are detected in total
F$_{\rm 850 \mu m}$,
F$^{\rm dust}_{\rm 850 \mu m}$  falls below this threshold; they
would not have been detected without the lift by the synchrotron contribution.
Although mainly characterised by upper limits, 
the averages of F$^{\rm dust}_{\rm 850 \mu m}$ are
2.53$\pm$2.45 mJy and 2.21$\pm$1.15 mJy
for the quasars and radio galaxies, resp.;
the higher rms for quasars may be caused by the 
subtraction of the larger synchrotron contribution. 
Thus, from both the number of sources with
F$^{\rm dust}_{\rm 850 \mu m}$\,$>$\,3\,mJy
and the mean fluxes, any 
F$^{\rm dust}_{\rm 850 \mu m}$  difference  
between quasars and radio galaxies is marginal.
Again, this is
consistent with the orientation-dependent unified schemes,
which predict for both AGN classes a similar 
dust power irrespective of aspect angle.

Actually, the 450 and 
850\,$\mu$m dust contribution of most sources is constrained only by
upper limits. A closer look at  Table\,1 of Willott et al. (2002) shows: 
While the 850\,$\mu$m rms is at least as good for the radio galaxy sample as for
quasars, at 450\,$\mu$m the rms of the radio
galaxies (17.3$\pm$6.8 mJy) is worse than that of the quasars (12.8$\pm$6.7 mJy),
indicating that the higher 450\,$\mu$m detection rate of the
quasars may be due to more favourable observing conditions. 
At 450\,$\mu$m only two quasars out of 11 are detected at the 3$\sigma$
level (3C191 and 4C35.23).
Subtracting the synchrotron extrapolation results in a
450\,$\mu$m dust contribution which is below the 3$\sigma$ level for all
except one quasar (3C191).
More sensitive FIR observations are required to
determine the dust properties of the two samples. As discussed
by Haas et al. (2003) for the radio-quiet quasars at z$\ga$1, the dust
temperature may be higher than the usually adopted 
T$\approx$50K, so that the emission will still peak at observed FIR
wavelengths and not in the sub-mm range.   

\section{Conclusions}
\label{section_conclusions}

MAMBO 1.25\,mm observations of nine 3CR quasars at z\,$\sim$\,1.5 yield the following results:
\begin{itemize}
\item[1)] For seven sources the 1.25\,mm data
  are much brighter than one would expect from a purely thermal dust model
  fitted to the submm data. This provides strong evidence that the 1.25\,mm fluxes 
  are dominated by synchrotron radiation. 
  Furthermore, in eight of nine cases the 1.25\,mm fluxes lie close to the power-law 
  extrapolation F$_\nu$$\propto$$\nu$$^{\alpha}$ of the total 
  synchrotron spectrum fitted at cm wavelengths, not indicating any pronounced  
  spectral curvature between 2\,cm and 1.25\,mm. 
\item[2)]  
  The lack of a strong spectral curvature between 
  1.25\,mm and 2\,cm, suggests that there is no 
  strong spectral break between 1.25\,mm and 850\,$\mu$m.
  If the synchrotron contribution at 850\,$\mu$m is consequently extrapolated from
  the 1.25\,mm data via a power-law F$_\nu$$\propto$$\nu$$^{\alpha}$, 
  then in at least 6 out of 9 
  quasars the observed 850\,$\mu$m fluxes are dominated by synchrotron
  radiation, too. In this case, only at most
  five quasars show evidence for substantial
  thermal 850\,$\mu$m emission.
\end{itemize}
In order to test the unified schemes, 
we compare the quasars with 12 radio galaxies at matched redshift and radio
power. 
As argued by Archibald et al. (2001) from the core spectra of some sources 
of their larger radio galaxy sample and indicated by our study of the total spectra, the 
850\,$\mu$m synchrotron contribution of the radio galaxies is low. This allows
for drawing the following conclusions:
%Although the radio galaxies do not yet have suitable 1.25\,mm data, the
%extrapolation of the power-law shaped synchrotron spectra, fitted at cm wavelengths, 
%provides upper limits to the synchrotron contribution at 850\,$\mu$m. 
%These upper limits allow for drawing the following conclusions: 
  
\begin{itemize}
\item[1)]  
  On average
  the synchrotron contribution at 850\,$\mu$m is
  by a factor of about 5
  higher in the quasars and the synchrotron spectral slope is 
  slightly steeper in the
  radio galaxies, as expected if the jet axis of the
  quasars is more aligned with our line of sight. 
\item[2)] 
  After subtraction of the synchrotron contribution we find evidence for
  thermal 850\,$\mu$m emission in two (possibly three) detected galaxies, and
  in three additional
  galaxies their high 850\,$\mu$m upper limits allow for dust emission, too. 
  Compared with the quasars, the distribution of the synchrotron subtracted 
  450 and 850\,$\mu$m dust emission shows no significant difference 
  to that of the radio
  galaxies, again consistent with the picture of a similar amount of
  isotropic dust power irrespective of aspect angle.
\end{itemize}
As a consequence, 
as long as future data do not reveal a synchrotron break between 850\,$\mu$m 
and 1.25\,mm for the quasars or a mm bump for the radio galaxies both
being not expected from the current data,
the data can be interpreted in accordance with the   
orientation-dependent unified schemes for powerful 
radio galaxies and quasars. In this case, 
our results challenge two speculative conclusions
drawn by Willott et al. (2002):
\begin{itemize}
\item[1)]  
  A luminosity dependent
  refinement of the unified schemes, the so-called receding torus model
  (Lawrence 1991),
  is neither required nor supported by the new data.
  As argued by Haas et al. (2004), at the high luminosities
  P$_{\rm 408 MHz}$ $\ga$ 10$^{\rm 24.5}$ W Hz$^{\rm -1}$
  the scale height of the torus might not be independent of luminosity,
  rather it may increase due to the
  impact of supernovae from starbursts accompanying the AGN phenomena. 
  This possibility is also considered
  by 
  Simpson (2005) who, however, still assumes that the 
  [O\,III]$_{\rm 500.7 nm}$ emission is isotropic. Since this
  assumption is not correct at high luminosities
  as shown by Haas et al. (2005) on the basis of {\it Spitzer} IRS
  spectra of 3CR
  sources
  %with the
  %{\it Spitzer Space Telescope}
  and the lower
  [O\,III]$_{\rm 500.7 nm}$\,/\,[O\,IV]$_{\rm 25.9 \mu m}$ ratio for
  galaxies compared with quasars,
  any remaining evidence for the simple version of the receding torus is further reduced. 
\item[2)]  
  The decline of dust luminosity with the projected linear
  size -- interpreted as age --  of the radio source (Fig.\,6
  in Willott et al. 2002) could be an artefact.
  Then the  trend in this figure might be caused by the
  higher 850\,$\mu$m synchrotron contribution in quasars. 
  Of course, evolution of the dust temperature and emission may exist, 
  as was concluded for the optically selected
  Palomar-Green quasars (Haas et al. 2003),
  but for the present samples of radio-loud quasars and radio
  galaxies at $z\sim1.5$ such evolutionary trends cannot yet be established 
  by the current submm data.
  Notably, Fanti et al. (2000) already found no evidence that the FIR 
  luminosities of
  the compact steep spectrum quasars and GHz peaked sources are
  significantly different from those of the extended objects.
\end{itemize}

The samples are still small, and the decomposition of the submm and mm data
into thermal, synchrotron (and free-free) components are affected by large
uncertainties, also due to varibility.
So far, however, the results from the 1.25\,mm data points strongly
indicate that Willott et al.'s  interpretation of the observed 
850\,$\mu$m difference between quasars and radio galaxies  
in favour of the receding torus or evolutionary trends
should be considered with reservation.

% ----------------------------------------------------------------
\acknowledgements 
%We wish to thank the anonymous referee for constructive suggestions. 
It is a pleasure for us to thank IRAM for discretionary observing time 
with the 30\,m telescope at Pico Veleta.
For literature and photometry search the NED %and SIMBAD
was used.
This work was supported by the 
Nordrhein-Westf\"alische Akademie der Wissenschaften.%,
%funded by the Federal State Nordrhein-Westfalen
%and the Federal Republic of Germany.
We thank the referee David Hughes for his detailed criticism, 
and Reinhard Schlickeiser for valuable discussions.


\begin{thebibliography}{}
%\bibitem{}   Akujor C.E., L\"udke E., Browne I.W.A., et al.,  1994, A\&AS  105, 247
%\bibitem{} Andreani P., Fosbury R.A.E., van Bemmel I., Freudling W., 2002 A\&A 381, 389
\bibitem{} Andreani P., Fosbury R., van Bemmel I., et al. 2002 A\&A 381, 389
%\bibitem{} Antonucci R.R.J., 1984, ApJ 278, 499
%\bibitem{} Antonucci R.R.J., Barvainis R., 1990, ApJL 363, L17
%\bibitem{} Antonucci R.R.J., Miller J.S., 1985, ApJ 297, 621
\bibitem{} Archibald E., Dunlop J., Hughes D.,  et al.,  2001, MNRAS 323, 417
\bibitem[Barthel (1989)]{Barthel89}     {Barthel P., 1989, ApJ 336, 606}
%\bibitem{} Barthel P.D., Arnaud K.A., 1996, MNRAS 283, L45
%\bibitem{}  Baum S.A., Zirbel E.L., O'Dea C.P., 1995, ApJ 451, 88
%\bibitem{} Bednarek W., Protheroe R.J., 1997, Proc. Rel. Jets in AGNs, 318
%\bibitem{} Begelman M.C., 1982, in IAU Symp. 97, eds.  Heeschen \& Wade, 223
%\bibitem{}   Begelman M.C., Blandford R.D., Rees M. J., 1984, RvMP 56, 255
%\bibitem{} Best P.N., R\"ottgering H.J.A., Bremer M.N., et al., 1998, MNRAS 301, L15
%\bibitem{} Bicknell G.V., 1994, ApJ 422, 542
%\bibitem{} Bicknell G.V., 1995, ApJS 101, 29
%\bibitem{} Bowman, M., Leahy J.P., Komissarov S.S., 1996, MNRAS  279, 899
%\bibitem{} Bregman J.N., Snider B.A., Grego L., Cox C.V., 1998, ApJ 499, 670
%\bibitem{} Cesarsky.C., Abergel A., Agnese P., et al., 1996, A\&A 315, L32
%\bibitem{} Chiaberge M., Capetti A., Celotti A., 1999, A\&A 349, 77
%\bibitem{} Chiaberge M., Capetti A., Celotti A., 2000, A\&A 355, 873
%\bibitem{} Chiaberge M., Gilli R., Macchetto F.D., Sparks W.B., Capetti A., 2003, ApJ 582, 645
%\bibitem{} Chini R., Kreysa E., Biermann P.L., 1989a, A\&A 219, 87
%\bibitem{} Chini R., Biermann P.L., Kreysa E., Gm\"und H.-P., 1989b, A\&A 221, L3
%\bibitem{} Cohen M.H., Ogle P.M., Tran H.D., Goodrich R.W., Miller
%J.S., 1999, AJ 118, 1963
\bibitem{} Crusius A., Schlickeiser R. 1986, A\&A 164, L16
%\bibitem{} de Koff S., Best P., Baum S., et al., 2000, ApJS 129, 33
%\bibitem{} de Vaucouleurs G., de Vaucouleurs A., Corwin jr. H.G., et
%  al. 1991, 'Third reference catalog of bright galaxies', Version 3.9,
%  Springer-Verlag, Berlin
%\bibitem{} de Young D.S., 1993, ApJL 405, L207
%\bibitem{}  Fanaroff, B. L.; Riley, J. M., 1974, MNRAS 167, 31
\bibitem[Fanti et al. (2000)]{Fanti2000} {Fanti C., Pozzi F., Fanti R., et al., 2000, A\&A 358, 499}
%\bibitem{} Freudling W., Siebenmorgen R., Haas M., 2003, ApJL 599, L13
%\bibitem{}  Golombek D., Miley G.K., Neugebauer G., 1988, AJ 95, 26
%\bibitem{} Gopal-Krishna,  Wiita P.J.,  2000, A\&A 363, 507
%\bibitem{} Gopal-Krishna,  Wiita P.J.,  2001, ASP Conf. Ser. 250, 290
%\bibitem{} Gopal-Krishna,  Wiita P.J.,  2001, A\&A 373, 100,/astro-ph0104326
%\bibitem{} Grimes J., Rawlings S., Willott C. 2004, MNRAS 349, 503
%\bibitem{} Haas M., 2001, Habilitationsschrift Universit\"at
%  Heidelberg, available from http://www.astro.ruhr-uni-bochum.de/haas/
%\bibitem{} Haas M., Chini R., Meisenheimer K., et al., 1998, ApJL 503, L109 
%\bibitem{} Haas M., M\"uller S.A.H., Chini R., et al., 2000, A\&A 354,  453
%\bibitem{} Haas M., Klaas U., M\"uller S.A.H., Chini R., Coulson I., 2001, A\&A 367, L9
\bibitem{} Haas M., Klaas U., M\"uller S., et al., 2003, A\&A 402, 87
\bibitem{} Haas M., M\"uller S., Bertoldi F., et al., 2004, A\&A 424, 531
\bibitem{} Haas M., Siebenmorgen R., Schulz B., Kr\"ugel E., Chini R.,
  A\&A submitted
%\bibitem{}   Heckman T.M., Smith E.P., Baum S.A., et al., 1986, ApJ 311, 526
%\bibitem[Heckman et al. (1992)]{Heckman92}
%  {Heckman T.M., Chambers K.C., Postman M., 1992, ApJ 391, 39}
%\bibitem[Heckman et al. (1994)]{Heckman94}
%       {Heckman T.M., O'Dea Ch.P., Baum S.A., Laurikainen E., 1994, ApJ 428, 65}
%\bibitem[Hes et al. (1995)]{Hes95}
%       {Hes R., Barthel P.D., Hoekstra H., 1995, A\&A 303, 8 }
%\bibitem{} Hildebrand R.H. 1983, QJRAS 24, 267
%\bibitem{}   Hill G.J., Goodrich R.W., Depoy D.L., 1996, ApJ 462, 163
%\bibitem{}  Hine R.G., Longair M.S., 1979, MNRAS 188, 111 
% \bibitem[Hoekstra et al. (1997)]{Hoekstra97}
%       {Hoekstra H., Barthel P.D., Hes R., 1997, A\&A 319, 757 }
%\bibitem{} Hughes D.H., Robson E.I., Dunlop J.S., Gear W.K., 1993, MNRAS 263, 607
%\bibitem{} Impey Ch.,  Gregorini L., 1993, AJ 105, 853
%\bibitem{} Jaffe W., Meisenheimer K., R\"ottgering H.J.A., et al. 2004, Nature 429, 47
%\bibitem{} Kellermann K.I., Sramek R., Schmidt M., et al., 1989, AJ 98, 1195
%\bibitem{} Kessler M.F., Steinz J.A., Anderegg M.E., et al., 1996, A\&A 315, L27
%\bibitem{} Kessler M.F., M\"uller T.G., Arviset C., Garc\'{\i}a-Lario  P., \& Prusti T. 2000,
%  {\it The ISO Handbook}, SAI/2000-035/Dc, ESA publ.
%\bibitem{} Kessler M., M\"uller Th.,  et al., 
%  {\it The ISO Handbook}, ESA-SP 1262
%\bibitem{} Klaas U., Laureijs R., Radovich M., Schulz B., Wilke K., 2002,
%  ``ISOPHOT calibration accuracies, V5.0 (adapted to OLP10)'' ISO Explanatory Library Doc. SAI/02-xxx/rp, ESA publ.
%\bibitem{} Klaas U., Haas M., M\"uller S.A.H., et al.,  2001, A\&A 379, 823
\bibitem{} Kreysa E., Gem\"und H.-P., Gromke J., et al., 1998, SPIE 3357, 319
%\bibitem{}   K\"uhr H., Witzel A., Pauliny-Toth I.I.K., Nauber U.,
%1981, A\&AS 45, 367
%\bibitem{}  Laing R.A. \& Bridle A.H., 2002, MNRAS 336, 1161
%\bibitem[Laing et al. (1983)]{Laing83}
%       {Laing R.A., Riley J.M., Longair M.S., 1983, MNRAS 204, 151}
%\bibitem{} Laing R.A., Parma P., de Ruiter H.R., et\,al., 1999, MNRAS 306, 513
%\bibitem{} Leeuw L., Sansom A., et al., 2004, ApJ  in press, astro-ph/0406011
%\bibitem{}  Laureijs R., Klaas U. 1999, ISOPHOT Error Budgets V1.0,
%  ISO Explanatory Library Doc. SAI/98-091/dc, ESA publ. 
%\bibitem{} Laureijs R., Klaas U., Richards P.J., Schulz B., Abraham
%  P., 2002,
%  The ISO Handbook, Vol. IV: PHT -- the Imaging Photo-Polarimeter, Version 2.0, SAI/99-069/dc, ESA publications
\bibitem{} Lawrence A., 1991, MNRAS 252, 586
%\bibitem{} Ledlow M.J., Owen F.N., 1996, AJ 112, 9
%\bibitem{}  Lemke D., Klaas U., Abolins J., et al.,  1996, A\&A 315, L64
%\bibitem{} Lilly S.J., Longair M.S., 1984, MNRAS 211, 833
%\bibitem{} Lilly S.J., Longair M.S., Miller L., 1985, MNRAS 214, 109
\bibitem{} Marscher A.P., 1977, ApJ 216, 244
%\bibitem{} Meier D.L., 1999, ApJ 522, 753
%\bibitem{} Martel A.R., Baum S.A., Sparks W.B., et al., 1999, ApJS 122, 81
\bibitem{} Meisenheimer K., Haas M., M\"uller S., et al.,  2001, A\&A 372, 719
%\bibitem{} Miley G., Neugebauer G., Soifer B.T., et al.,  1984, ApJL 278, L79
%\bibitem{} Neugebauer G., Green R.F., Matthews K., et al., 1987 ApJS 63, 615
\bibitem{} Orr M., Browne I., 1982, MNRAS 200, 1067
\bibitem{} Pacholzyk A., 1970, {\it Radio astrophysics}, W.H. Freeman
  \& Comp. San Francisco
%\bibitem{} Owen F., Ledlow M., 1994, ASP Conf. Ser. 54, 319
%\bibitem{} Pier E.A., Krolik J.H., 1992, ApJ 401, 99
%\bibitem{} Pier E.A., Krolik J.H., 1993, ApJ 418, 673
%\bibitem{} Polletta M., Courvoisier Th., Hooper E.J., Wilkes B.J., 2000, A\&A 362, 75
%\bibitem{}  Robson E.I., Leeuw L.L. Stevens J.A., Holland W.S., 1998, MNRAS 301, 935
\bibitem{} Simpson C. 2005, MNRAS 360, 565   
%\bibitem{} Simpson C., Ward M., Wall J.V., 2000, MNRAS 319, 963
%\bibitem{} Singal, A.K., 1996, MNRAS 278, 1069
%\bibitem{} Siebenmorgen R., Freudling W., et al., 2004, A\&A 421, 129
%\bibitem{} Snellen I., Schilizzi R., de Bruyn A., Miley G., 1998, A\&A 333, 70
%\bibitem{} Spinrad H., Djorgowski S., Marr J., Aguilar L., 1985, PASP 97, 932
%\bibitem{} Steppe H., Salter C.J., Chini R., et al., 1988, A\&AS 75, 317
%\bibitem{} Steppe H., Liechti S., Mauersberger R., et al., 1992, A\&AS 96, 441
%\bibitem{} Steppe H., Paubert G., Sievers A., et al., 1993, A\&AS 102, 611
%\bibitem{} Stevens J. A., Robson E. I., Gear W. K., et al.,  1998, ApJ 502,
%\bibitem{} Stickel M., Lemke D., Klaas U., et al.,  2000, A\&A 359, 865
%\bibitem{} Urry C.M., Padovani P., 1995, PASP 107, 803
%\bibitem{} Valtonen M.J., Hein\"am\"aki P., 2000, ApJ 530, 107
%\bibitem[van Bemmel et al.(2000)]{vanBemmel2000} {van Bemmel I.M., Barthel P., de Graauw Th., 2000,  A\&A 359, 523}
\bibitem[van Bemmel et al.(2001)]{vanBemmel2001} {van Bemmel I., Bertoldi F., 2001,  A\&A 368, 414}
%\bibitem{} Verdoes Kleijn G., Baum S., de Zeeuw T., et al., 2002, AJ 123,1334
%\bibitem{} Whysong D., Antonucci R., 2003, NewAR 47, 219
%\bibitem{} Whysong D., Antonucci R., 2004, ApJ 602, 116
%\bibitem{} Willott C.J., Rawlings S., Blundell K.M., Lacy M.., 2000, MNRAS 316, 449
%\bibitem{} Willott C.J., Rawlings S., Archibald E.N., Dunlop J.S., 2002, MNRAS 331, 435
\bibitem{} Willott C., Rawlings S., Archibald E., et al. 2002, MNRAS 331, 435
%\bibitem{} Zirbel E.L., Baum S.A., 1995, ApJ 448, 521
\end{thebibliography}
\end{document}